\documentclass[conference]{IEEEtran}
\usepackage[utf8]{inputenc}
\usepackage{color,xcolor}
\usepackage{epsfig}
\usepackage{amsmath,amsfonts,amsthm,amssymb}
\usepackage{graphics,graphicx,pgfplots,tikz}
\usepackage{float}
\usepackage{algorithm,algorithmic}
\usepackage{svg}
\usepackage{epstopdf}
\usepackage{soul,csquotes,ulem}
\usepackage[hidelinks]{hyperref}
\usepackage[font=footnotesize]{subcaption}
\usepackage[font=footnotesize]{caption}

\usepackage{balance}

\begin{document}
\title{Fusion of Time and Angle Measurements for Digital-Twin-Aided Probabilistic 3D Positioning}
	\author{
		\IEEEauthorblockN{
			Vincent Corlay, Viet-Hoa Nguyen, and Nicolas Gresset
		}
		\thanks{
			The authors are with Mitsubishi Electric Research and Development Centre Europe, 35700 Rennes, France (e-mail: \{v.corlay, v.nguyen, n.gresset\}@fr.merce.mee.com).
		}
	}
	\maketitle
	

\begin{abstract}
Previous studies explained how the 2D positioning problem in indoor non line-of-sight environments can be addressed using 
ray tracing with noisy angle of arrival (AoA) measurements.
In this work, we generalize these results on two aspects. 
First, we outline how to adapt the proposed methods to address the 3D positioning problem.
Second, we introduce efficient algorithms for data fusion, where propagation-time or relative propagation-time measurements (obtained via e.g., the time difference of arrival) are used in addition to AoA measurements.
Simulation results are provided to illustrate the advantages of the approach.
\end{abstract}


\section{Introduction}
Positioning is a crucial aspect of modern wireless communication networks. It enables various applications, such as indoor navigation and location-based services. 
However, positioning based on wireless systems faces many challenges including multipath propagation and interference, while a high accuracy is often required.
Emerging technologies like 5G and 6G are expected to enhance the positioning capabilities.

In line-of-sight (LoS) situations, standard positioning algorithms such as triangulation and trilateration can be applied.  
However, this paper targets indoor positioning.
Indoor environments are challenging due to obstacles such as walls and clutters, which can block signals or cause reflections, resulting in a non-line-of-sight (NLoS) scenario.
Standard positioning algorithms therefore suffer from unreliability and inaccuracy due to the unavailability of LoS conditions. 
To overcome these challenges, a 3D model of the environment and ray-tracing simulations can be used to recover the NLoS paths of the radio signal.

Several types of measurements are commonly used for positioning, namely time of departure (ToD), time of arrival (ToA), time of flight (ToF), time difference of arrival (TDoA), angle of arrival (AoA), received signal strength indicator (RSSI), etc.
Each measurement has its own advantages and drawbacks, making it suitable for specific scenarios. 
Data fusion positioning systems that combine multiple measurements can significantly improve overall performance and should therefore be considered for challenging indoor environments.
  
\textbf{Related work by the same authors.}
In previous works \cite{CorlayA}\cite{CorlayB}, we explained how ray tracing can improve indoor positioning using the ``reverse ray-tracing" approach.
More specifically, we considered the case of noisy angle of arrival measurements.  
Given the uplink received signal from the user equipment (UE) to be located, we proposed to launch rays, from several base stations (BS), in a Monte Carlo manner according the measurement error statistics \cite{CorlayA}. 
For the 2D positioning problem, this generates a set of points in a 2D plane for each BS: The locations where the launched rays cross the 2D plane of interest. 
Then, we proposed to fit a 2D distribution on the set of obtained points of each BS \cite{CorlayB}. 
Finally, multiplying the fitted distributions provides a position probability at each candidate location. 
This improves robustness with fewer launched rays when done online or allows all ray launching steps and distribution fitting operations to be performed offline such that the probability-distribution parameters are directly recovered from a storage table for online inference \cite{CorlayB}.  

\textbf{Related work by other authors.} 
References \cite{Kaya2007}\cite{Kong2006}\cite{Kong2016}\cite{Ryzhov2023}\cite{Voltz1994} explore a similar framework that employs reverse ray tracing for positioning based on AoA measurements and a digital twin of the environment.
Nevertheless, the probabilistic Monte-Carlo approach is not investigated to mitigate the measurement noise.
The readers are invited to consult \cite{CorlayA}\cite{CorlayB} to obtain more details about these studies.

\textbf{Main contributions.} In this paper, we extend our previous works \cite{CorlayA}\cite{CorlayB} on two important points:
\begin{itemize}
\item  We explain how the proposed approach can be extended
to the 3D positioning case.
\item For accuracy improvement, data fusion is investigated.  We explain how time-related additional measurements, such as ToA, ToD, or TDoA, can be utilized to improve the positioning accuracy.
\end{itemize}


\section{Considered framework and notations}

\subsection{Problem statement and baseline solution}

\begin{figure}[t]
    \centering
    \includegraphics[scale=0.58, trim =0cm 0cm 0cm 0cm, clip]{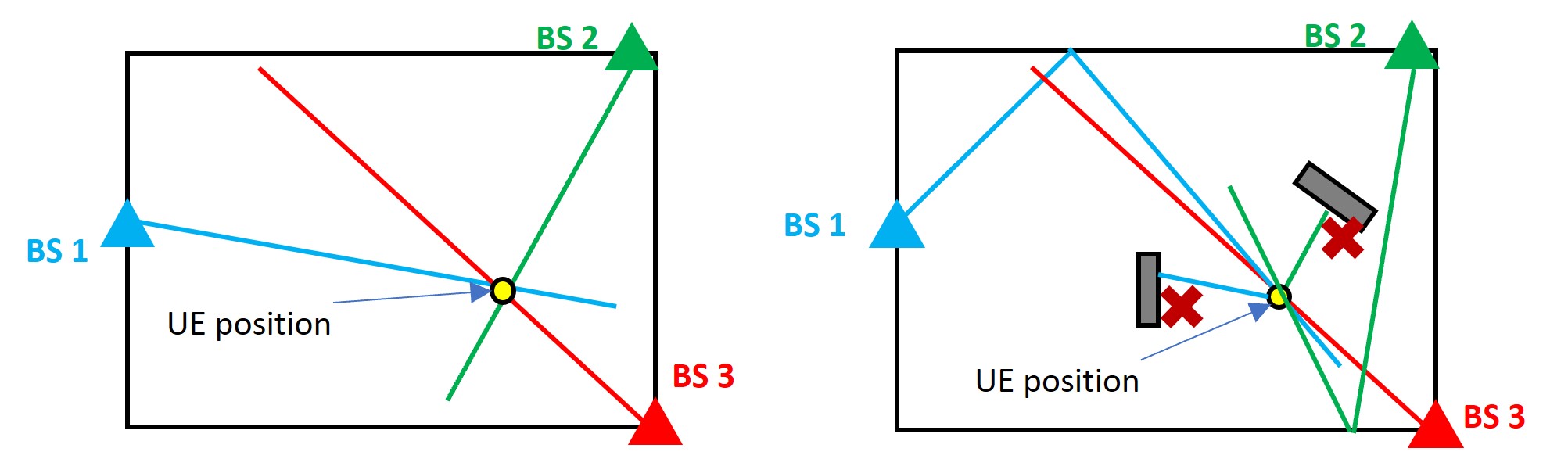}
	\caption{Left: Standard triangulation approach via  LoS-based AoA at the BS. Right: Situation where some LoS AoA are not available for several BS. In the example, two of the three LoS paths are blocked by clutters. In this case,  reverse ray tracing using a digital twin can be considered. }
    \label{fig:example_triang_reverse_ray}
\end{figure}

This paper considers the following digital twin-based technique to address the NLoS uplink positioning problem. 
Several BS perform AoA measurements on an uplink signal transmitted by a UE. 
Then, using a digital twin of the environment, ray tracing in these AoA directions is performed. 
The intersection of the rays is then the estimated position of the UE. This approach, illustrated in Figure~\ref{fig:example_triang_reverse_ray}, can be called “reverse/backward ray tracing”.

When dealing with noisy AoA measurements, we can follow this procedure: For each BS, we sample the AoD for the rays to be launched according to the statistical characteristics of the AoA measurement error. 
We refer to this method as the Monte Carlo approach.
Thus, numerous rays are launched from each BS. The result of this ray-launching procedure is a collection of points in the xy plane at the elevation of the UE, indicating where the rays intersect the xy plane.
The distribution fitting approach, explained in the following section, is then used to obtain the probability distributions for each BS. 
Finally, these distributions are to be multiplied to obtain the position probabilities.

The angle information can be obtained by processing the received signal via several signal-processing algorithms such as  Delay-and-Sum \cite{DS}, MUSIC \cite{MUSIC}, MVDR \cite{MVDR}, ESPRIT \cite{ESPRIT}. 
Each algorithm may lead to a different error statistics.

\subsection{Ray-tracing propagation model}

A ray-tracing propagation model is considered to compute the path of the radio waves in the environment. 
It enables to compute the candidate trajectories between a transmitter and a receiver, even in NLoS situations.

In this paper, a ray has either an infinite length (i.e., case of infinite number of bounces) or a finite length. 
In the following sections, we propose to discretize the rays into points. Then, each point as an associated length corresponding to the length between the start of the ray and the point position.
The term ``length of a ray" refers to the last point of a discretized ray if the ray length is finite.

Finally, regarding the propagation of the rays, we consider only specular reflections. Please consult \cite[Sec. II]{CorlayB} for a discussion on this model assumption.

\subsection{Statistical modeling and notations}

Let $X$ be a random variable representing the position of a UE. Let $\theta$ be a random variable representing the true AoA of the signal, and $Y$ a random variable representing the BS measurement(s) of the AoA.
Regarding the notations, we use $p(\theta|y)$ for $p(\Theta = \theta|Y=y)$ and similarly $p(x|y)$ for $p(X=x|Y=y)$. 
The distribution $p(\theta|y)$ represents the statistics of the uplink AoA measurement error.

We let $n$ be the number of measurements such that $\mathbf{y}=[y_1,y_2,…,y_i,…,y_n ]$.  
The vector $\mathbf{y}$ comprises the measurement performed by all BS. For the sake of simplicity, we assume that there is one set of measurements per BS, and therefore $n$ BS. The measurement $y_i=[y_i^1,y_i^2,...]^T$ may contain measurements of several categories, such as the AoA and ToA (see below). 

In the considered positioning problem, the goal is to approximate $p(x|\mathbf{y})$.

\section{3D positioning}

This section explains how  the proposed approach of \cite{CorlayA}\cite{CorlayB} can be extended
to the 3D positioning case. 

\subsection{Obtaining 3D points}
\begin{figure}[t]
\vspace{-13mm}
    \centering
    \includegraphics[scale=0.59]{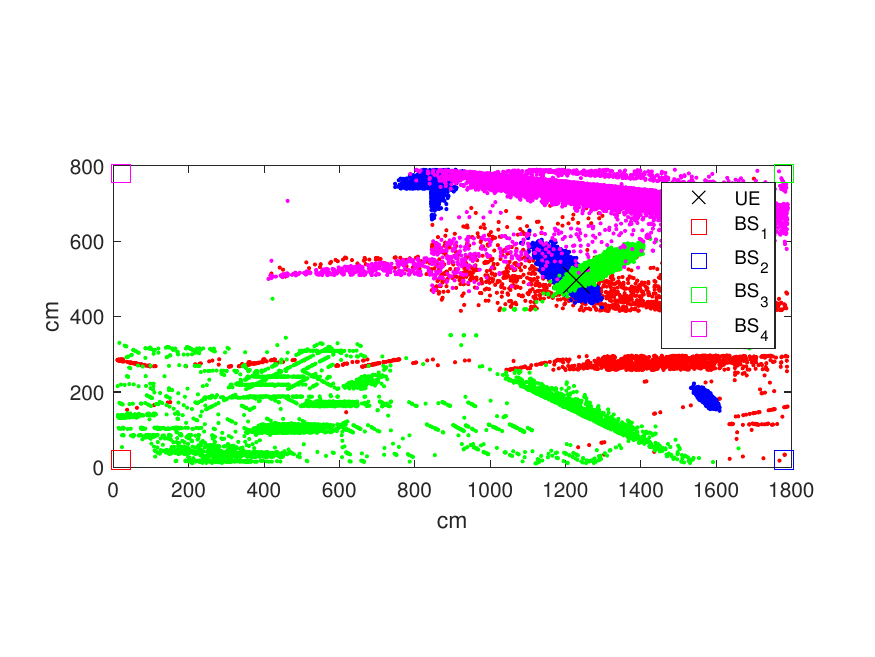}
\vspace{-14mm}
    \caption{(2D positioning problem) Map of points corresponding to the positions where the rays launched by each BS cross the xy plane. }
    \label{fig:map_points}
\vspace{-2mm}
\end{figure}

For the 2D positioning case, points are obtained as the intersection between the launched rays and the 2D plane of interest. The elevation of this plane is the (assumed to be) known elevation coordinate $z$. This is illustrated in Figure~\ref{fig:map_points}. There is one colour for the set of points obtained for each BS.

\begin{figure}[t]
\centering
    \includegraphics[width=0.8\linewidth, trim = 5.5cm 1cm 5cm 1cm, clip]{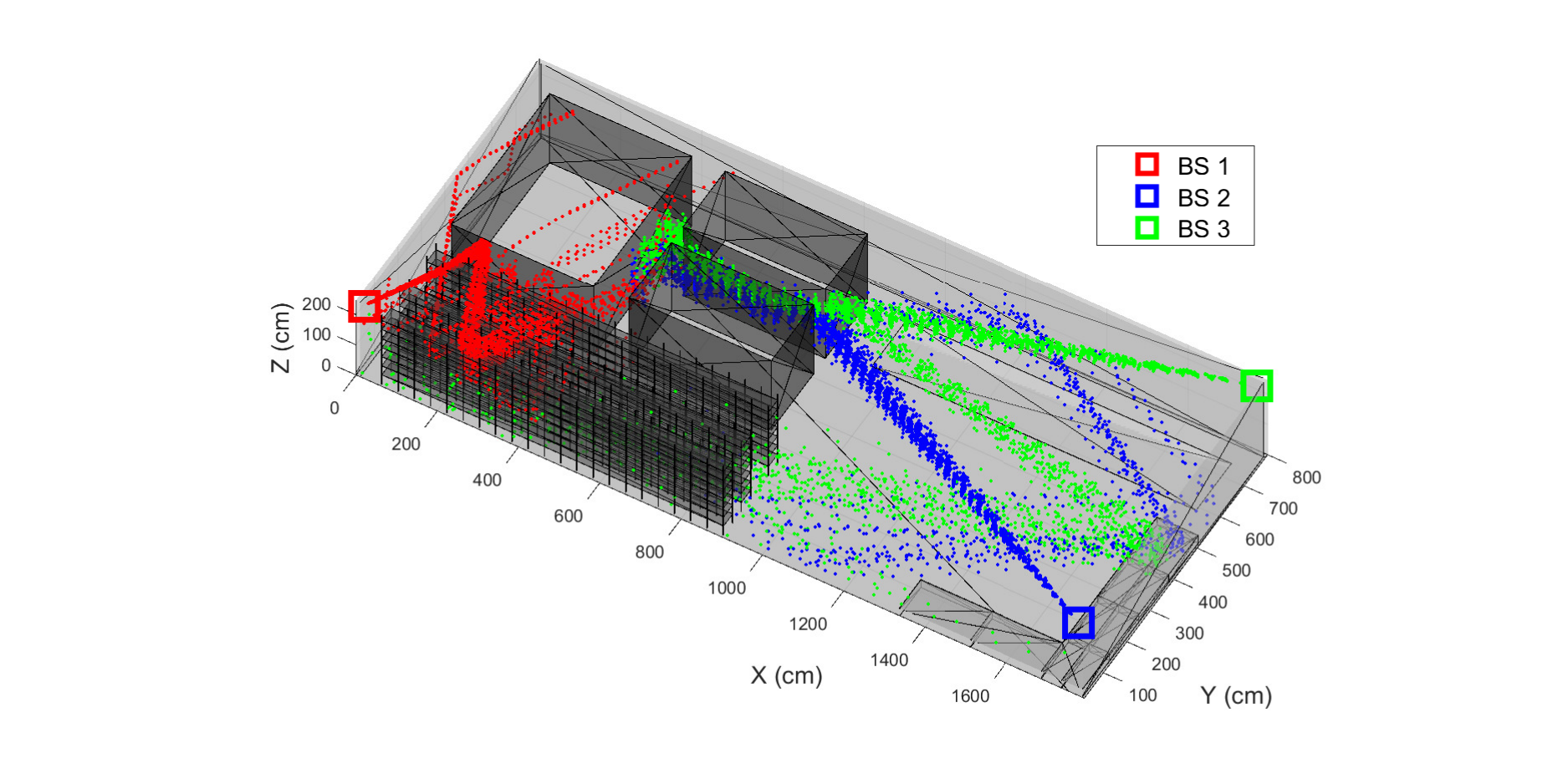}
    \caption{Set of 3D points obtained by discretizing the rays. Theses points are obtained using AoA measurements only.}
    \label{fig:points_AoA}
\end{figure}

For the 3D positioning approach, we suggest discretizing the rays into points through sampling.
This yields 3D points as shown in Figure~\ref{fig:points_AoA}.
In this example, the angle of the rays are chosen in a Monte Carlo manner based on noisy AoA measurements and their error statistics. 

\subsection{Fitting 3D probability distributions and multiplying them}
\label{fitt_3D}

\begin{figure}[t]
    \centering
    \includegraphics[scale=0.59]{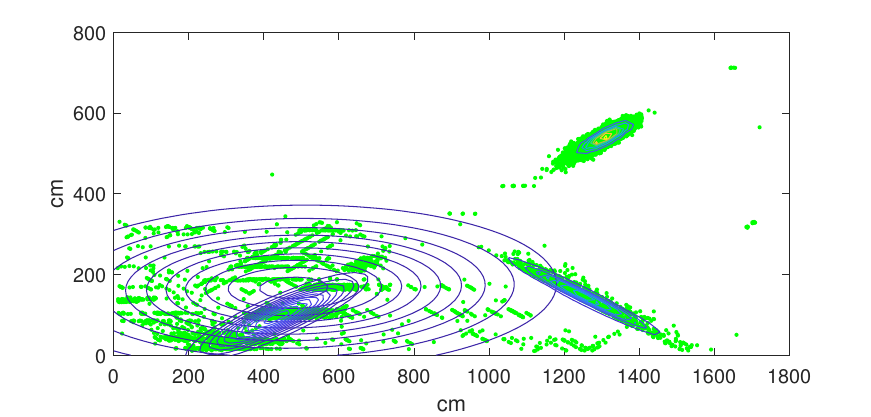}
\vspace{-2mm}
    \caption{(2D positioning problem) Green: Map of points obtained when launching the rays from one BS with a given AoA and a given error statistics. Blue: Contour plot of the clusters of the fitted GMM.}
    \label{fig:fitted_GMM}
\end{figure}

For the 2D case, for each set of points obtained from each BS (one color in Figure~\ref{fig:map_points}) a parametric distribution is fitted.
The distribution can for instance be a 2D Gaussian mixture model (GMM), as illustrated in Figure~\ref{fig:fitted_GMM}.

\begin{figure}[t]
\centering
    \includegraphics[width=0.8\linewidth, trim = 3cm 1cm 3cm 1cm, clip]{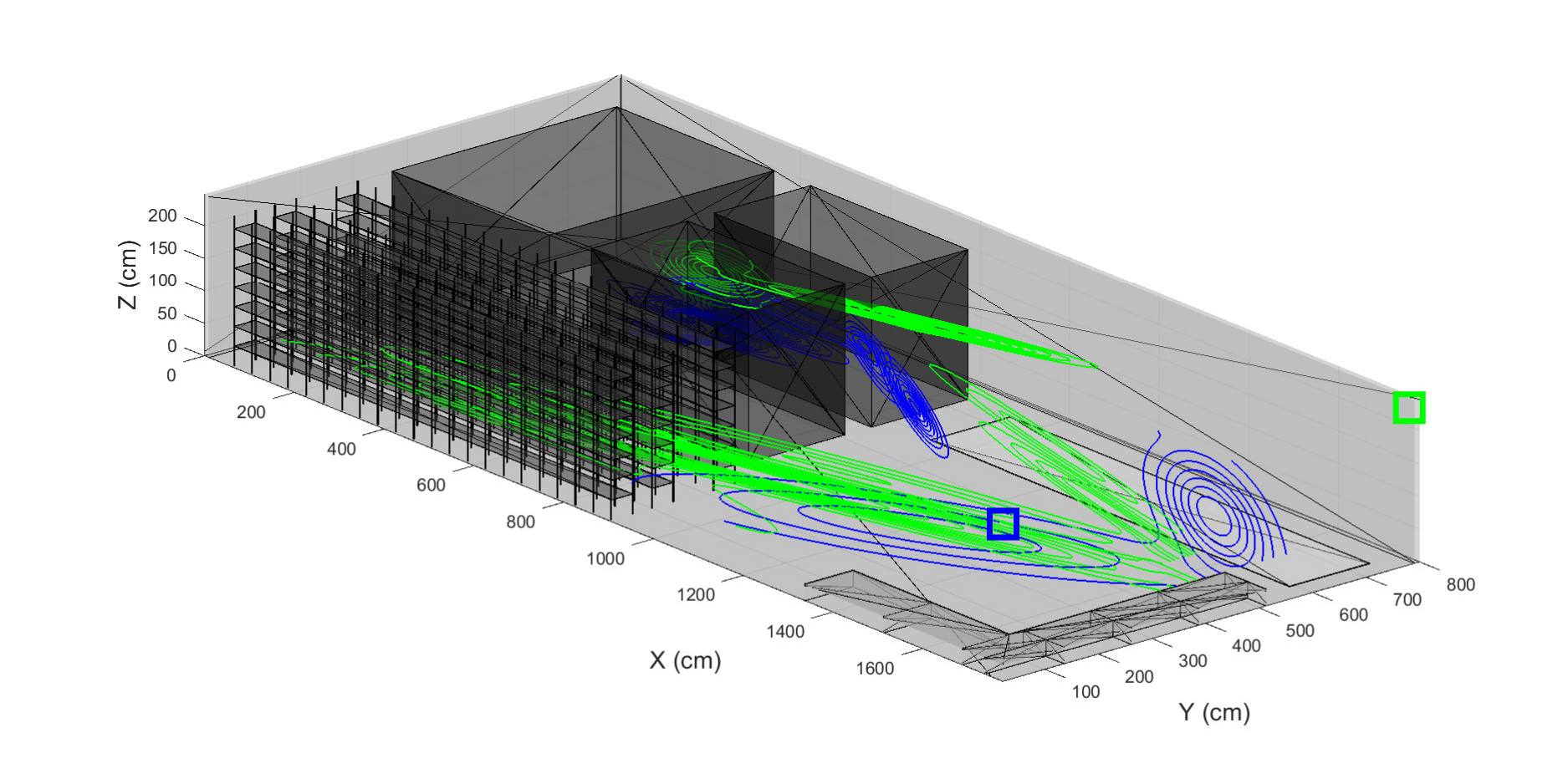}
    \caption{Contour plot of obtained 3D GMM probability distributions using the points of Figure~\ref{fig:points_AoA} (based on AoA measurements only). }
    \label{fig:gmm_AoA}
\end{figure}

For the 3D case, the same approach can be considered.  
Instead of using a 2D probability distribution, a 3D distribution should be used. 
One can fit 3D GMM on the set of 3D points obtained for each BS. 
As in the 2D case, the position probabilities are then obtained by multiplying the obtained 3D probability density functions. 
The same fitting algorithm as in \cite{CorlayB} (expectation-maximization) can be used to compute the parameters of the probability density functions. This is illustrated for the 3D case in Figure~\ref{fig:gmm_AoA}. 
Simulation results provided at the end of the paper indicate that the 3D GMM distribution is adapted for the considered set of 3D points.

\section{Propagation time and relative propagation time}

In this section, we present the two categories of measurements considered for data fusion with the AoA measurements. 

\subsection{Propagation time}

In addition to AoA measurements, a BS may also have information related to a propagation time (PT) of the signal between the UE to localize and the given BS. 
This PT information can be obtained in several manners such as:
\begin{itemize}
\item By computing a difference between a ToA and ToD of the signal to obtain the ToF. One important constraint when measuring the PT in this manner is to have a fine clock synchronization between the UE and the BS.
\item By considering the delay corresponding to the largest coefficients of the channel impulse response.
\item By analyzing the RSSI.
\end{itemize}

Let us denote a PT measurement at BS $i$ as $y^1_i$. 
This PT information can be mapped to a suited length of ray $L$ or equivalently to the length associated to a 3D point. One should simply use the formula:
$L=y^1_i\cdot c$, where $c$ is the propagation speed of the signal, such as the speed of light.

Given a probability distribution on the PT, we can deduce a probability distribution $p(L|y^1_i)$ on the length of the points or on the length of the rays (if it is finite). 
This distribution is analogous to the distribution $p(\theta|y^2_i)$ (where $y^2_i$ is an AoA measurement). 

\subsection{Relative propagation time}
\label{sec_RPT}

In some cases, a relative propagation time (RPT) rather than a PT may be available. 
This RPT information can be obtained from a time TDoA of the signal between two BS. 
It is therefore computed as the difference of two measured ToA at two different BS. 
Similarly to the PT, the RPT can be mapped to the distance information by using the constant $c$.

The main advantage of the RPT is that no synchronization between the UE to localize and the BS is required. 
Only the BS should be synchronized.
As the BS are usually higher-quality hardware and connected, this is a high advantage.

The RPT provides information about the length difference between two points corresponding to two BS (points of different colors in Figure~\ref{fig:points_AoA}), but not on the length of the points. 
An example of point-length difference is shown in Table~\ref{fig:example_RPT}.
In the example, there are 4 point lengths for the first BS (obtained from one or more rays launched from this BS) and 5 point lengths for the second BS (obtained from rays launched from this second BS). This results in $4\times 5=20$ possible pairs. Those values are to be compared with an expected point-length difference $\Delta L$ obtained from the RPT measurement of the two considered BS. 

\begin{table}[h]
    \centering
    \includegraphics[scale=0.80]{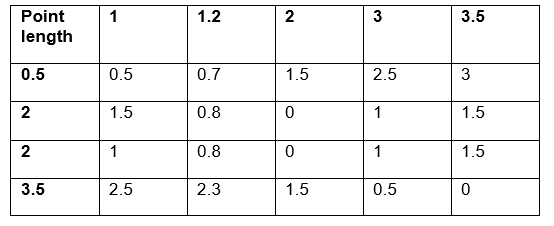}
    \caption{In the first row: the lengths of 5 points obtained via ray launching from a first BS. In the first column: the lengths of the 4 points obtained via ray launching from a second BS. The table values represent the length difference between the various points. }
    \label{fig:example_RPT}
\end{table}

Given an error statistics on the RPT, one has access to the distribution $p(\Delta L|y_i^1,y_j^1)$, where $y_i^1,y_j^1$ represents the ToA of the signal at BS 1 and BS 2, used to compute the TDoA and therefore the RPT and length difference.
As a result, the points of Figure~\ref{fig:points_AoA} can be discriminated based on to their length difference:
\begin{itemize}
\item Pairs of points having a likely length difference should be kept.
\item Pair of points having unlikely length difference should be discarded.
\end{itemize}

We explain how this operation can be realized in the next section.

\section{Point selection with propagation time and relative propagation time}
This section explains how additional measurements, PT or RPT, can enhance the positioning accuracy obtained with AoA measurements only.

\subsection{Fusion of angle of arrival and propagation time measurements}
\label{sec_selec_PT}
Each point associated to a ray has the same probability given an AoA measurement but not given a PT measurement. 
Hence, an infinite-length ray is not associated to a probability distribution on the PT $p(L|y)$, but rather the points of the subsampled rays. 
As a result, to sample the points according to $p(L, \theta |y_i^1,y_i^2)$, one could proceed as described in the following \textbf{Algorithm 1}  (Monte Carlo approach). \\

\begin{algorithm}
\caption{Point selection based on PT and AoA measurements}
\begin{algorithmic}[1]
\FOR {each BS $i$}
\STATE Launch the rays. Either $(i)$ in all directions (case no AoA measurement is available) or $(ii)$ according to $p(\theta|y^2_i)$ if AoA measurements are available.
\STATE Sort the points, obtained by discretizing the rays, by length and make bins of small size.
\STATE Sample several distances according to the PT distribution $p(L|y_i^1)$. Each distance falls within one bin. Choose randomly a point in this bin.
\ENDFOR
\end{algorithmic}
\end{algorithm}

Alternatively, one can switch step 3 and 4:  For a given ray, one can first sample a distance according to $p(L|y_i^1)$ and keep only the point having this distance on the launched ray.

This process enables to obtain, for each BS $i$, a set of points sampled according to $p(L, \theta |y_i^1,y_i^2)$.

\subsection{Fusion of angle of arrival and relative propagation time measurements}
\label{sec_selec_RPT}
 As explained in Subsection~\ref{sec_RPT}, one aims at selecting pairs of points having a likely length difference.
One should therefore first collect the length difference associated to all pairs of points, sort them by length, and select pairs of points by Monte Carlo sampling.
The following \textbf{Algorithm 2} describes this process. It is similar to Algorithm 1, but where the RPT is used instead of the PT.\\

\begin{algorithm}
\caption{Point selection based on RPT and AoA measurements}
\begin{algorithmic}[1]
\FOR {any pair of 2 BS $i,j$}
\STATE Launch the rays. Either $(i)$ in all directions (case no AoA measurement is available) or $(ii)$ according to $p(\theta|y^2_i)$ if AoA measurements are available.
\STATE Compute the length difference of any pair of points from the two different BS. Sort the points by length difference and make bins of small size.
\STATE Sample several length difference according to the RPT distribution $p(\Delta L|y_i^1,y_j^1)$. Each length difference falls within one bin. Choose randomly a pair of points in this bin.
\ENDFOR
\end{algorithmic}
\end{algorithm}

This process enables to obtain, for each BS $i$, a set of points sampled according to the probability distribution $p(\Delta L, \theta |y_i^1,y_j^1,y_i^2)$.

\subsection{Obtained subset of points and distribution fitting}

Both approaches described in Subsection~\ref{sec_selec_PT} and \ref{sec_selec_RPT} enables to down-select the points of Figure~\ref{fig:points_AoA}, which are selected only based on AoA measurements. 
For instance, using PT measurements for the down-selection leads to the subset of points shown on Figure~\ref{fig:points_AoA_PT}.

Subsequently, this subset of points is to be used for distribution fitting described in Section~\ref{fitt_3D}. 
This yields the fitted distributions (one distribution per BS) illustrated in Figure~\ref{fig:gmm_AoA_PT}, which are more relevant compared to ones in Figure~\ref{fig:gmm_AoA}.

As a reminder, the position-probability estimates of the UE are then obtained by multiplying the obtained fitted distributions.

\section{Performance evaluation}
\subsection{Error model}
For AoA, PT, and RPT measurements, a Gaussian error model is considered:
$p(\theta|y^2) \sim N(\theta , \sigma_{\eta}^2)$, $p(L|y^1) \sim N(L, \sigma_{\nu}^2)$, and $p(\Delta L|y_i^1,y_j^1) \sim N(\Delta L, \sigma_{\nu}^2)$.

\subsection{Simulation environment}

\begin{figure}[t]
    \centering
    \includegraphics[scale=0.3]{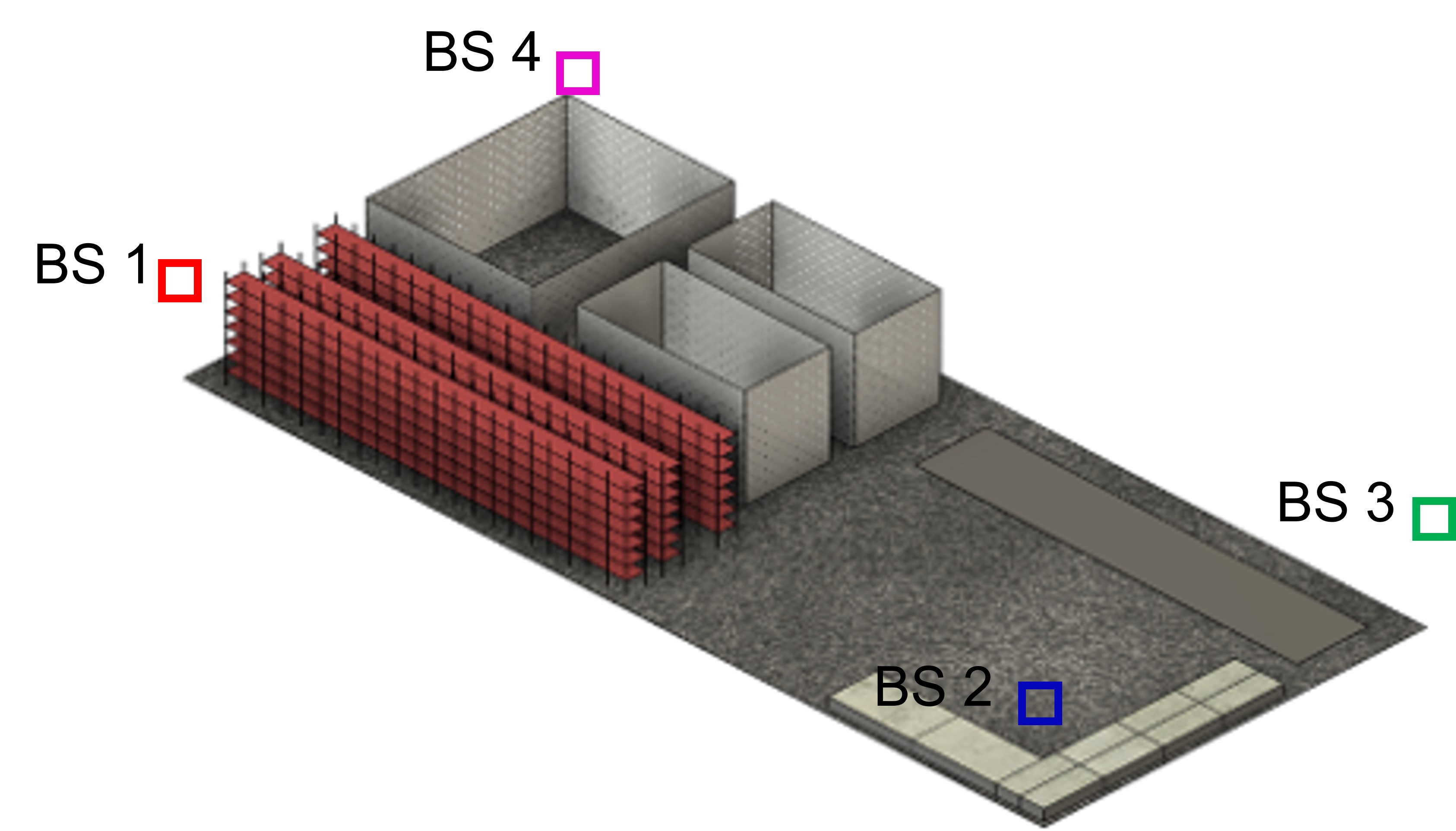}
    \caption{Considered environment for the simulations: 4 BS are located in the top corners, illustrated by 4 different colors}
    \label{fig:scene_digi-twin}
\end{figure}

We use the same simulation environment as in \cite{CorlayA}\cite{CorlayB}.
The scene is illustrated in Figure~\ref{fig:scene_digi-twin} and has the following dimensions: width 8m, length 18m, and height 2.5m. 
This environment is inspired from the recommendations of the 5G Alliance for Connected Industries and Automation (5G-ACIA) for indoor industrial scenario \cite{5G_ACIA}.
The 4 BS are located in the top corners of the scene (see Figure \ref{fig:scene_digi-twin}). The UE is randomly dropped in the considered environment (i.e, with a uniform distribution). Hence, there is randomness both in the UE position and the BS measurement errors. 

The positioning error is denoted by $\epsilon = ||\hat{x} - x||$, where $\hat{x}$ corresponds to the position with the highest probability according to distribution obtained as the multiplication of the fitted GMM models and $x$ is the ground-truth position.

\subsection{Results}

\begin{figure}[!h]
\centering
    \includegraphics[width=0.8\linewidth, trim = 5.5cm 1cm 5cm 1cm, clip]{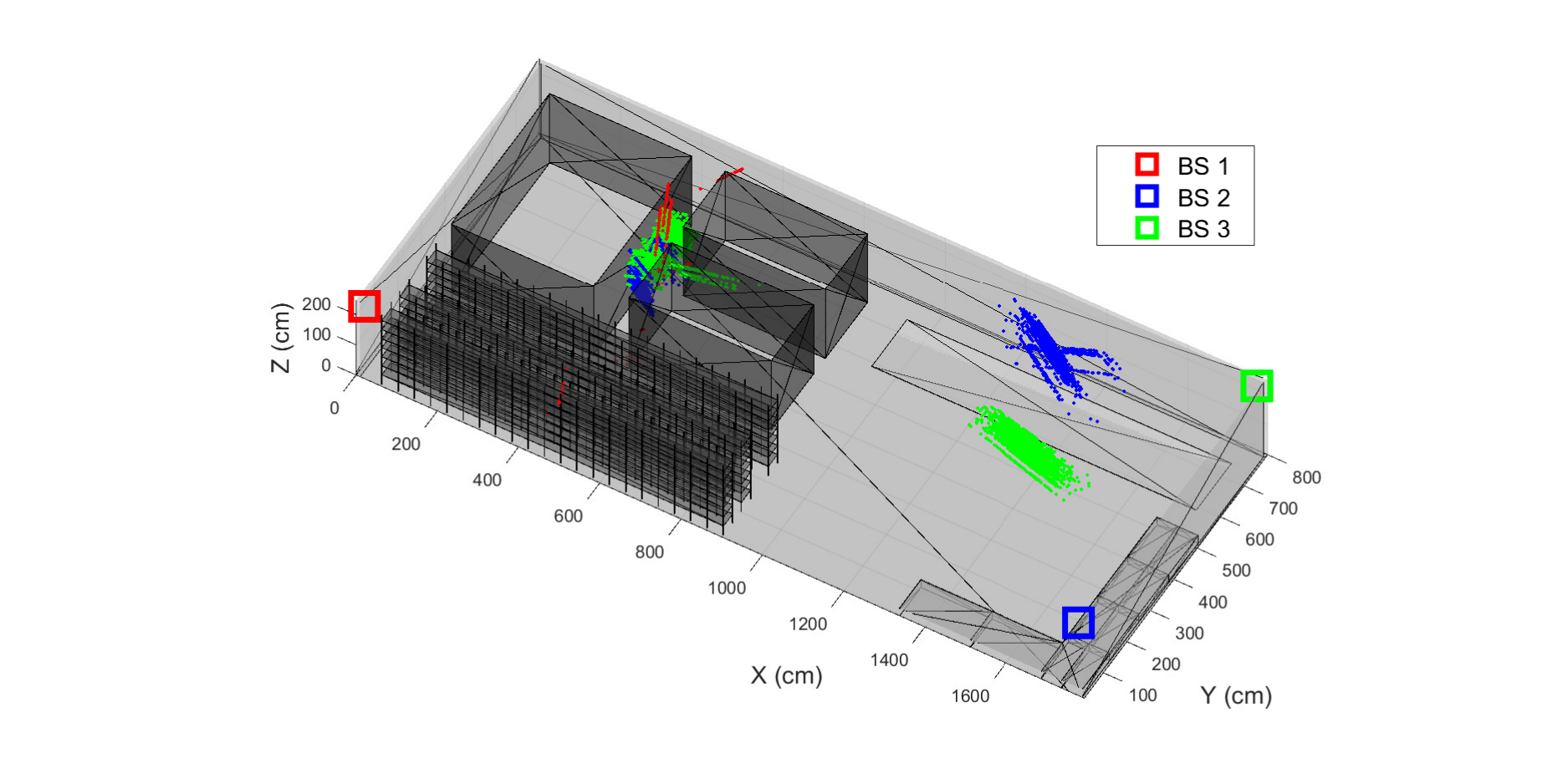}
    \caption{Obtained points in 3D using AoA and PT measurements.}
    \label{fig:points_AoA_PT}
\end{figure}

\begin{figure}[!h]
\centering
    \includegraphics[width=0.8\linewidth, trim = 3cm 1cm 3cm 1cm, clip]{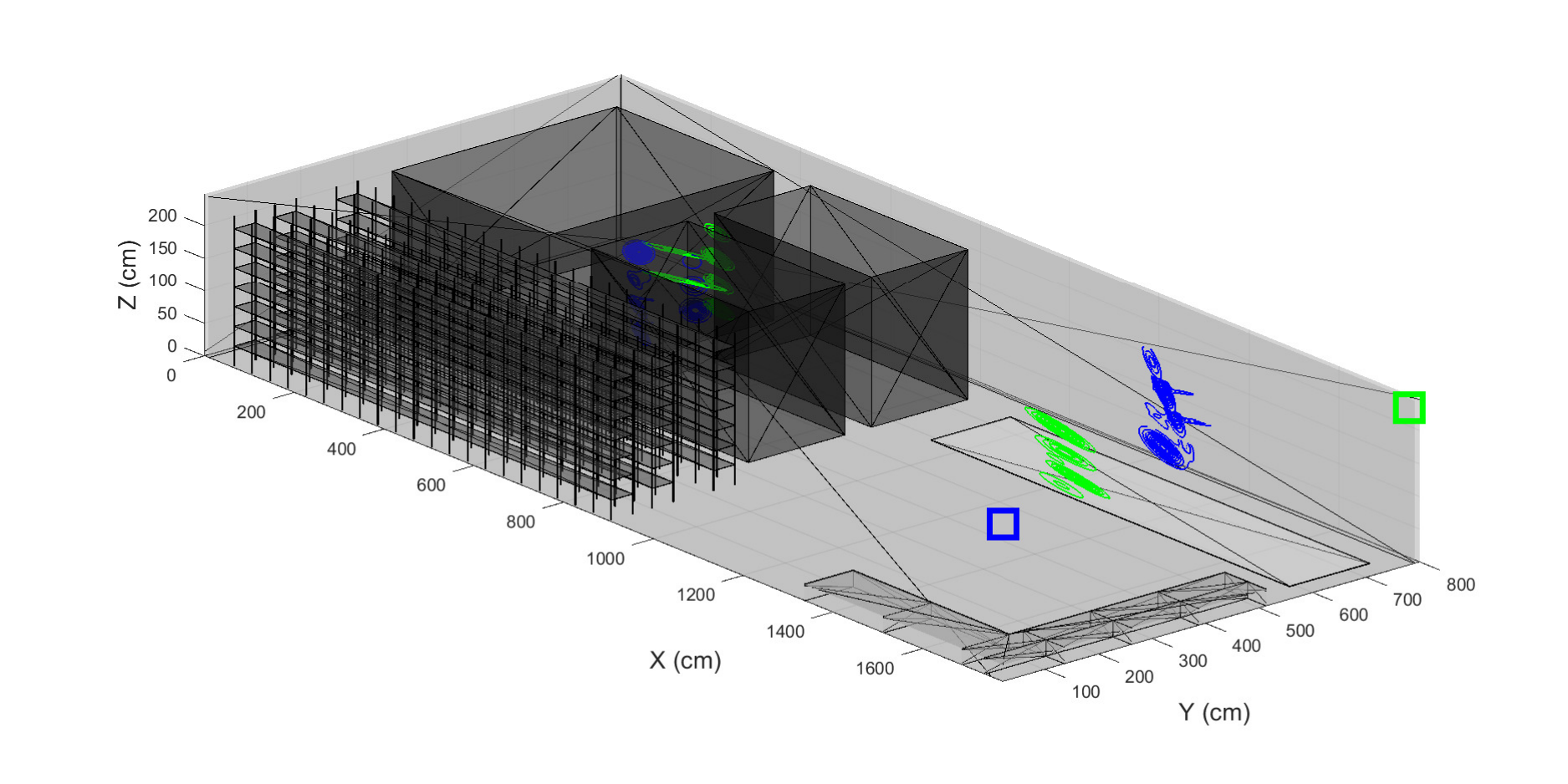}
    \caption{Contour plot of 3D GMM probability distributions using both  AoA and PT measurements.}
    \label{fig:gmm_AoA_PT}
\end{figure}

Figure \ref{fig:points_AoA} and \ref{fig:points_AoA_PT} show the points obtained from using only AoA measurements and both AoA and PT measurements, respectively. 
When both AoA and PT measurements are used, the obtained points concentrate into a reduced number of zones. 
With AoA measurements only, the points are distributed along the propagation paths.  
The set of possible positioning solutions is therefore reduced with the additional PT measurements. This phenomenon is shown in Figure  \ref{fig:gmm_AoA} and \ref{fig:gmm_AoA_PT}.

\begin{figure}[!h]
\centering
    \includegraphics[width=0.94\linewidth]{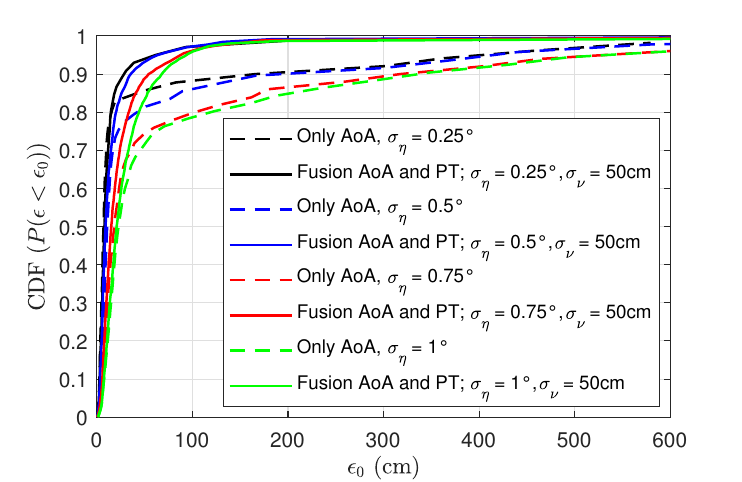}
    \caption{Cumulative density function of the positioning error without and with PT measurements (both in 3D).}
    \label{fig:Simu_AoAvsAoAPT}
\end{figure}

In Figure \ref{fig:Simu_AoAvsAoAPT}, the cumulative distribution function (CDF) of the positioning error is presented for cases both without and with PT measurements.
The set of considered standard deviation of AoA noise is $\{$0.25°, 0.5°, 0.75°, 1°$\}$ and the standard deviation of the PT noise is 50 cm. 
We observe that the fusion of AoA and PT measurements significantly improves the positioning accuracy. 
For example, with a standard deviation of the AoA noise of 1°, adding PT measurement can lower the positioning error of 250 cm at the 90th percentile.

Figure \ref{fig:Simu_AoAvsAoAPT} also shows the effect of increasing the AoA noise. 
The positioning error increases as the AoA noise becomes more significant.
For example, with $\sigma_{\eta}=$0.25°, the 90th percentile is below 28 cm. With 1° this number increases to 67 cm. 
\begin{figure}[!h]
\centering
    \includegraphics[width=0.94\linewidth]{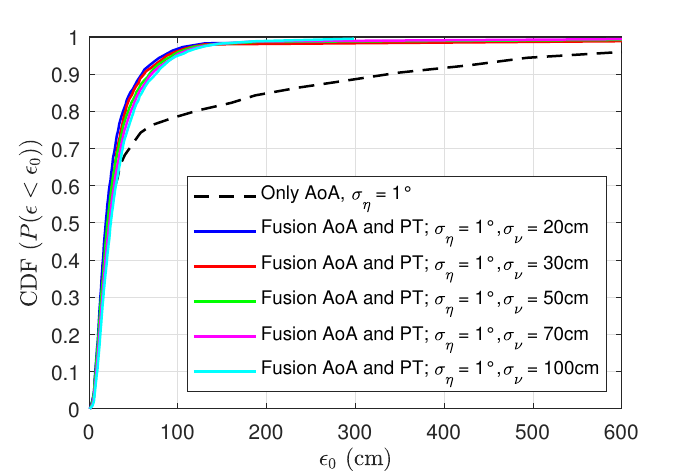}
    \caption{Cumulative density function of the positioning error with different standard deviations of the PT noise.}
    \label{fig:Simu_AoA1_PTall}
\end{figure}

In Figure \ref{fig:Simu_AoA1_PTall}, we report the CDF of the positioning error when the AoA noise standard deviation is 1° and when the PT noise standard deviation ranges in $\{20, 30, 50 ,70 , 100\}$ cm. 
As expected, reducing the noise of the PT measurements improves positioning accuracy. However, the impact of the PT noise is significantly reduced compared to the AoA noise. %
This can be explained as follows: Using AoA measurements only, multiple candidate positions are determined as the intersections of the likely modes of the GMM distributions. 
The PT information then helps distinguish between likely and unlikely intersections. 
As the number of such intersections is limited and may be widely spaced, even with high variance in the PT information, the candidate intersections can still be effectively differentiated.

\begin{figure}[!h]
\centering
    \includegraphics[width=0.94\linewidth]{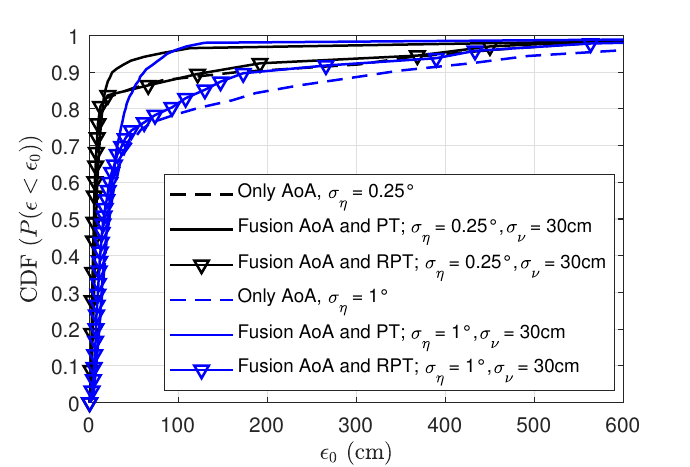}
    \caption{Cumulative density function of the positioning error with different standard deviations of the RPT noise.}
    \label{fig:Simu_AoA_PT_RPT_2}
\end{figure}

Finally, Figure \ref{fig:Simu_AoA_PT_RPT_2} reports simulation results when RPT measurements, instead of PT measurements, are combined with AoA. 
The performance improvement is less than that of the PT with the same standard deviation.
This is not unexpected as the RPT measurements offers less direct information than the PT measurements. 
However, it is important to note that, from a system perspective, the RPT information is easier to acquire than the PT information.
Nevertheless, using these additional RPT measurements enables better positioning performance than with only AoA measurements when the AoA measurement noise is high (1°).

\section{Conclusions}
This paper proposes a novel technique that combines ray-tracing-based methods with data fusion of angle-based and propagation-time-based measurements to enhance the positioning accuracy in indoor NLoS scenarios. 
Moreover, while previous studies focused on the 2D positioning problem, we explained how to utilize existing principles to enable 3D positioning.
Simulation results demonstrate significant improvements in positioning performance with data fusion compared to using angle measurements only.

\balance


\end{document}